\documentclass[%
 reprint,
nofootinbib,
 amsmath,amssymb,
 aps,
]{revtex4-2}

\usepackage{graphicx}
\usepackage{dcolumn}
\usepackage{bm}
\usepackage{hyperref}

\begin{document}

\preprint{APS/123-QED}

\title{Search of the pair echo signatures in the high-energy light curve of GRB190114C}

\author{Ievgen Vovk}
 \email{vovk@icrr.u-tokyo.ac.jp}
\affiliation{%
 Institute for Cosmic Ray Research, The University of Tokyo\\
 5-1-5 Kashiwa-no-Ha, Kashiwa City, Chiba, 277-8582, Japan
}%


%

\date{\today}

\begin{abstract}
A model of the time delayed electromagnetic cascade ``echo'' is applied to the bright gamma-ray burst GRB190114C -- the first gamma-ray burst to be contemporaneously detected in high and very high energy gamma-ray bands. It is shown that the internal spread of the cascade in the absence of the intervening magnetic fields dilutes the ``echo'' emission over $10^3-10^5$~seconds depending on the energy.
Accounting for the measured source flux in the $0.3-1$~TeV gamma-ray band, the prediction of the ``echo'' model is shown to match the detected lower-energy gamma-ray emission $10^4$~seconds after the burst.
However, the ``echo'' emission remains indistinguishable from the intrinsic GRB190114C flux within the measurement uncertainties.
Implications of this in the context of the intergalactic magnetic field measurement are discussed.
\end{abstract}

\maketitle


\section{Introduction}

Propagation of the very-high-energy (VHE; $\gtrsim100$~GeV) gamma rays over cosmological distances inevitably leads to their partial absorption due to interaction with the extragalactic microwave and infrared/optical photon fields~\cite{gould66, jelley66, stecker92}. The power absorbed is transferred to the electromagnetic cascades initiated in the process, eventually transforming the initial multi-TeV photons into lower, GeV energy secondary gamma-ray ``pair echo'' emission~\cite{aharonian_coppi, plaga95, neronov_semikoz_09}. Development of these cascades, spanning over $\sim 100$~Mpc distances, is sensitive to the intervening magnetic field, providing a potential opportunity to measure the extremely weak magnetic field in the intergalactic space~\cite{neronov_semikoz_09,neronov10a,durrer13}.
In the presence of the non-negligible intergalactic magnetic field (IGMF), the cascade and the corresponding secondary gamma ray emission are spread both in time and in angle, reducing the observable secondary flux. Consequently, non-detection of such secondary emission from several hard-spectrum blazars has been used to set a lower limit on the strength of IGMF~\cite{neronov10, taylor11, tavecchio11, vovk12, fermi_igmf, HEGRA_IGMF, MAGIC_extended, HESS_IGMF, Veritas_IGMF} at redshift $z \sim 0.1$. 
The nature of this IGMF remains, however, uncertain.

An opportunity to distinguish the astrophysical and cosmological origins of IGMF would open if its redshift evolution is traced to the redshifts of $z \sim 1-2$, where the field of an astrophysical origin should not have yet developed~\cite{garaldi21}.
Obtaining IGMF constraints at such redshifts is, however, challenging due to the progressively increasing with redshift gamma-ray absorption, limiting the number of detectable persistent gamma ray emitters.
Several transient sources, though -- such as gamma-ray bursts (GRB) and flaring active galactic nuclei (AGN) -- have been reported detected at TeV energies at redshifts $z \sim 0.5 - 1$~\cite{magic_pks1222, magic_pks1441, magic_b0218, magic_grb190114c_a}, potentially allowing to expand the redshift range of IGMF measurements / constraints.

Indeed, HE (high-energy; $\gtrsim 100$~MeV) and VHE observations of such transients have been discussed earlier as a viable tool to constrain IGMF~\cite{razzaque04, ichiki08, murase08, takahashi08, murase09}.
Detectability of the secondary emission from such transients, required for IGMF measurement, depends crucially on the intrinsic time delay of the cascade emission, caused by its internal scatter due to the angular spreads of the electron-positron pair production and consequent inverse Compton (IC) emission. 
Such effects can be accounted for with the recently developed general-purpose Monte Carlo (MC) codes ELMAG, CRPopa and CRBeam~\cite{ELMAG, CRPropa, CRbeam}, where the time delay can estimated as a difference between the propagation duration of the source primary and cascade secondary photons. However, the required accuracy of $\epsilon \simeq c \Delta t / d \sim 10^{-17}$ (for $\Delta t = 10$~s and $d\approx 10^{28}$~cm at redshift of $z=1$) may be challenging to achieve with them\footnote{E.g. the commonly used double-precision floating-point format has the precision of $\epsilon = 2^{-53} \approx 10^{-16}$~\cite{ieee_754}}, calling for specialized calculations of cascade light curves of transients sources.
Semi-analytical models, neglecting the exact angular dependencies of these processes, predict that at GeV energies this time delay may be as large as $\Delta t \sim 10^2-10^4$~s~\citep[e.g.][]{takahashi08, neronov_semikoz_09}, potentially exceeding the duration of GRBs or even short AGN flares~\cite[e.g.][]{hess_pks2155, magic_ic310}.

Perhaps, the first opportunity to search for such cascade emission at redshifts larger than $z \sim 0.1$ has presented itself with contemporaneous HE and VHE observations of a bright GRB190114C at redshift $z \approx 0.42$ with Fermi/LAT and MAGIC telescopes~\cite{magic_grb190114c_a, magic_grb190114c_b}. In this manuscript the measured VHE light curve of GRB190114C is used to predict the lower energy cascade emission, that is then tested against the HE measurements. To this end, a refined description of the cascade time delayed light curve, accounting for the exact energy and angular dependencies of pair production and IC emission, is developed and presented below.

\section{Calculation of the pair echo emission}
\label{sect::pair-echo-calculation}

Here it is assumed that the cascade starts with the absorption of the initial VHE gamma ray in interaction with the extragalactic background light photon field (EBL), modelled following~\cite{franceschini08}. Electrons and positrons, born in the process, emit secondary photons in IC scattering of the cosmic microwave background (CMB) (given that CMB photon number density exceeds that of EBL by more than two orders of magnitude, additional IC scattering on EBL has a minor effect). Only the first generation of the pairs and their IC emission is included in the calculations. For the initial source spectra not exceeding few TeVs (case for the measured VHE emission of GRB190114C), secondary gamma rays of the first generation do not exceed few tens of GeV in energy. The emission from the subsequent generations thus falls to MeV energies, outside of the energy range Fermi/LAT~\cite{FermiLAT}.


In the absence of IGMF, the angular distribution of the pairs, responsible for internal cascade scatter, is set by the corresponding pair production cross section, a back-reaction from inverse Compton emission~\cite[{e.g.}][]{ichiki08} and the angular profile of the initial VHE emission. However, for TeV energies considered here the pairs angular spread due to IC back-reaction is 
$\sqrt{\left<\theta_{IC}\right>} \approx 2 \times 10^{-8} \gamma_6^{-1/2}$~rad~\cite{ichiki08} 
(where $\gamma_6$ is the electron/positron Lorentz factor in units of $10^6$) -- much smaller than the typical $\sim 10^{-6} \gamma_6^{-1}$~rad spread of the pair production or IC emission itself and may be neglected. Therefore only angular (and energy) distributions of the pair production and IC emission are considered here.

With a single cascade generation considered, the calculation of the secondary emission is performed in the hybrid semi-analytical / Monte Carlo manner in the following steps: (a) calculation of the radial (redshift) distribution of the $e^+/e^-$ pairs injected following the initial VHE photons absorption, (b) calculation of their angular and energy distributions, (c) random sampling of the pairs following these distributions, (d) calculation of the secondary gamma-ray emission via IC scattering, accounting for the corresponding particle energy loss and (e) calculation of the time delay for every point of the electron / positron trajectories. These are outlined below.

\subsection{Radial distribution of the electron-positron pairs}

Radial distribution of the electron-positron pairs resulting from the initial gamma rays of energy $E_\gamma$ emitted at the redshift $z_e$ is set by the optical depth $\tau(E_\gamma, z_e)$ to the absorption of such photons on the extragalactic background light, that in the differential form can be written as~\cite{franceschini08}
\begin{eqnarray}
  \frac{d\tau(E_\gamma, \mu)} {d\mu dz} = 
  c \frac{1 - \mu}{2} \frac{dt}{dz}
  \int^{\infty}_{\frac{2 m_e^2 c^4}{E_\gamma \epsilon (1-\mu) (1 + z)}}
  d\epsilon 
  \frac{dn(\epsilon, z)}{d\epsilon}
  \sigma_{\gamma\gamma}
  \label{eq::dtau-dz-ds}
\end{eqnarray}
where $\mu = \cos \theta$ is the cosine of the angle between the interacting gamma ray and background photons. The cumulative probability of the electron-positron pair creation by the redshift $z$ is thus simply
\begin{equation}
  P(z, z_e) = e^{-\int_{-1}^1 d\mu \int_z^{z_e} dz \frac{d\tau(E_\gamma, \mu)} {d\mu dz}}
  \label{eq::ebl-probability-vs-redshift}
\end{equation}
which gives the required distribution.

However, since the energy and angular distribution of the generated pairs depends on $\mu$ (see below), it is convenient to consider the differential distribution $P(z, z_e, \mu)$ integrated only over $z$ when generating the pairs in the Monte Carlo approach employed here.

\subsection{Angular dependence of the photon-photon pair production}
\label{sect::pp-angular}

Differential cross section of the photon-photon pair production for two photons with energies $E_{\gamma}$ and $\epsilon$ may be written as~\cite{lee98}
\begin{eqnarray}
  \frac{d\sigma_{\gamma\gamma}}{dx} = 
    \sigma_T
    \frac{3}{4}
    \frac{m_e^2 c^4}{s}
    \left[
      \frac{x}{1-x}
      +
      \frac{1-x}{x}
        + \frac{1 - \beta_{cm}^2}{x (1-x)}
    \right. \nonumber \\
    \left.
      - \frac{\left(1-\beta_{cm}^2\right)^2}{4 x^2 (1-x)^2}
    \right ]
    \label{eq::pp-dsigma-de}
\end{eqnarray}
where $x = E_e / E_\gamma$ and $E_e$ is the generated electron (or positron) energy, $s = 2 E_{\gamma} \epsilon (1 - \cos \theta)$
is the squared center of momentum energy and $\beta_{cm} = \sqrt{1 - 4 m_e^2 c^4 / s}$ is the velocity of the resulting electron in the CM reference frame. The range of $x$ is restricted to $(1 - \beta_{cm}) / 2 \le x \le (1 + \beta_{cm}) / 2$.

As both generated electron and positron in the CM frame have equal oppositely directed velocities, their angular distribution may be obtained from the Lorentz transformations between the laboratory and CM frames following
~\cite{svensson82, coppi_blandford90}
\begin{equation}
  \mu = \frac{\mu' + \beta_c} {1 + \beta_c \mu'} 
    = \frac
      {\frac{2 x \omega_1}{\omega_1 + \omega_2} + \beta_c^2 \beta_{cm} - 1}
      {\beta_c \left(\frac{2 x \omega_1}{\omega_1 + \omega_2} + \beta_{cm} - 1\right)}
  \label{eq::pp-mu}
\end{equation}
where 
$\mu = \cos \theta$ is the cosine of the electron (positron) movement direction with respect to the direction of the CM frame movement in the laboratory frame, 
$\mu' = \left(2 \gamma / (\omega_1 + \omega_2) - 1 \right) / \beta_c \beta_{cm}$ is that in the CM frame,
$\gamma = E_e / m_e c^2 = x \omega_1$ is the electron (positron) Lorentz factor in the laboratory frame,
$\gamma_{cm} = \sqrt{s / 4 m_e^2 c^4}$ -- that in the CM frame, $\beta_c = \sqrt{1 - 4 \gamma_{cm}^2 / (\omega_1 + \omega_2)^2}$ is the velocity of the CM frame in the laboratory one with the additional notations $\omega_1 = E_{\gamma} / m_e c^2$ and $\omega_2 = \epsilon / m_e c^2$ introduced for compactness.
The pair production differential cross section as a function of the generation angle can be then expressed as
\begin{eqnarray}
  \frac{d\sigma_{\gamma\gamma}}{d\mu} = 
    \frac{d\sigma_{\gamma\gamma}}{dx} \frac{dx}{d\mu} \frac{d\mu}{d\mu'}
    \nonumber \\
    =
    \frac{d\sigma_{\gamma\gamma}}{dx} 
    \frac{\beta_c}{2 (1 - \beta_c^2)}
    \frac
    {[2 x \omega_1 + (\beta_{cm} - 1) (\omega_1 + \omega_2)]^2}
    {\beta_{cm} \omega_1 (\omega_1 + \omega_2)}
  \label{eq::pp-dsigma-dmu}
\end{eqnarray}

\begin{figure}
  \includegraphics[width=\columnwidth]{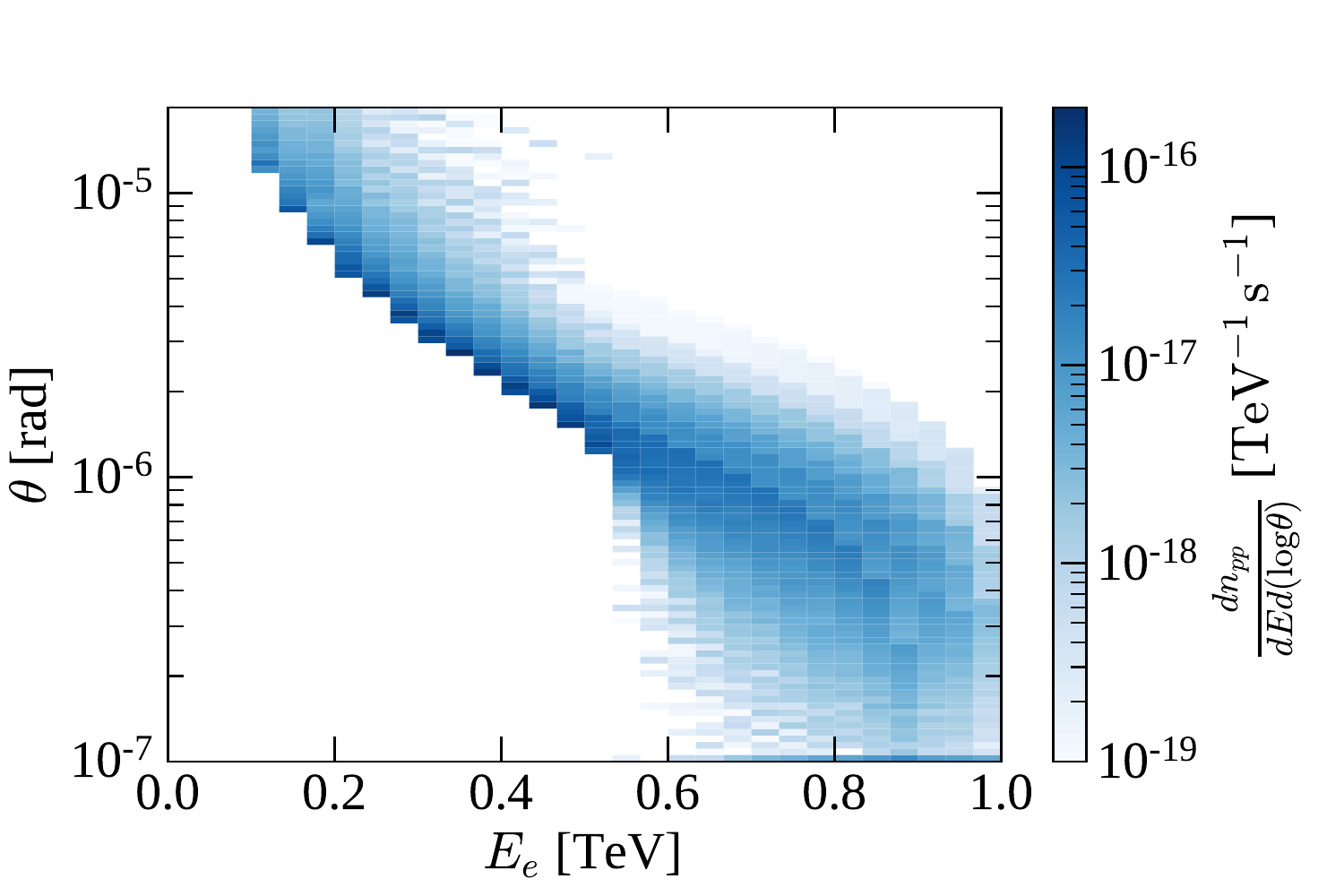}
  \caption{
    Differential pair production rate for $E_{\gamma} = 1$~TeV gamma ray as function of the generated electron (positron) energy and motion direction offset angle with respect to that of the gamma ray. Calculations performed for the EBL photon field at redshift $z=0$. Apparent layering at $E_e \gtrsim 0.6$~TeV is a numerical artifact coming from the energy binning of the used EBL model~\cite{franceschini08}.
  }
  \label{fig::pp-rate-example}
\end{figure}

For an isotropic target photon distribution the cross section dependency on electron energy can be found integrating Eq.~\ref{eq::pp-dsigma-dmu} over all the incident angles $\theta$. An example of such integration for the case $E_{\gamma}=1$~TeV gamma ray interacting with EBL at redshift $z=0$ is shown in Fig.~\ref{fig::pp-rate-example}. One can see that for most of the generated particle energy range, the particles are injected at the hollow cone with the $\theta \sim 2/\gamma$ opening with respect to the CM frame direction as result of the both the particles and the CM frame motions.

The CM frame movement direction in general does not coincide with that of the most energetic of the interacting photons. Assuming $\omega_1 \gg \omega_2$ and accounting for the pair production threshold of $\omega_1 \omega_2 \ge 1$, the corresponding maximal offset angle can be estimated as $\theta_{CM} \approx 2 \omega_2 / \omega_1 \le 2 \omega_2^2$. For the cosmic microwave radiation and extragalactic background light target fields with $\omega_2 \lesssim 1$~eV the CM frame offset is $\theta_{CM} \lesssim 10^{-12}$~rad, which is several orders of magnitude smaller than the typical $1/\gamma \gtrsim 10^{-7}$~rad direction spread of the $E_e<10$~TeV electrons, considered below. Due to this, the corresponding coordinate frame rotation is neglected here.

\subsection{Angular dependence of the Inverse Compton emission}

Inverse Compton emissivity for a single electron or positron on a monochromatic background photon distribution with energy $\epsilon_0$ may be written following~\cite{brunetti00} as
\begin{eqnarray}
  j(\Omega_{sc}, \epsilon_1) = 
    \frac{r_0^2 c n}{2 \gamma^2}
    \frac{\epsilon_1}{L_1} \delta (\epsilon - \epsilon_0) 
    \left[
      \left(
        1 + \frac{\epsilon_1}{m c^2} \frac{{\bf km} - 1}{\gamma L}
      \right)^{-1} 
      \right. \nonumber \\
      \left.
      + \frac{\epsilon_1}{m c^2} \frac{{\bf km} - 1}{\gamma L}
      + \left(
        1 + \frac{{\bf km} - 1}{\gamma^2 L L_1}
      \right)^2
    \right]
    \label{eq::ic-emissivity}
\end{eqnarray}
with $L = 1 - \beta {\bf ke}$ and $L_1 = 1 - \beta {\bf me}$ where $\bf e$, $\bf k$ and $\bf m$ are the direction vectors of the electron, incoming and scattered photons correspondingly. Directions of the incoming background photons that may be scattered from the energy $\epsilon$ up to $\epsilon_1$, are given by the relation
\begin{eqnarray}
  {\bf km} = 1 + \frac{m c^2}{\epsilon \epsilon_1}
    \left(
      \epsilon_1 \gamma L_1 - \epsilon \gamma L
    \right)
\end{eqnarray}

\begin{figure}
  \includegraphics[width=\columnwidth]{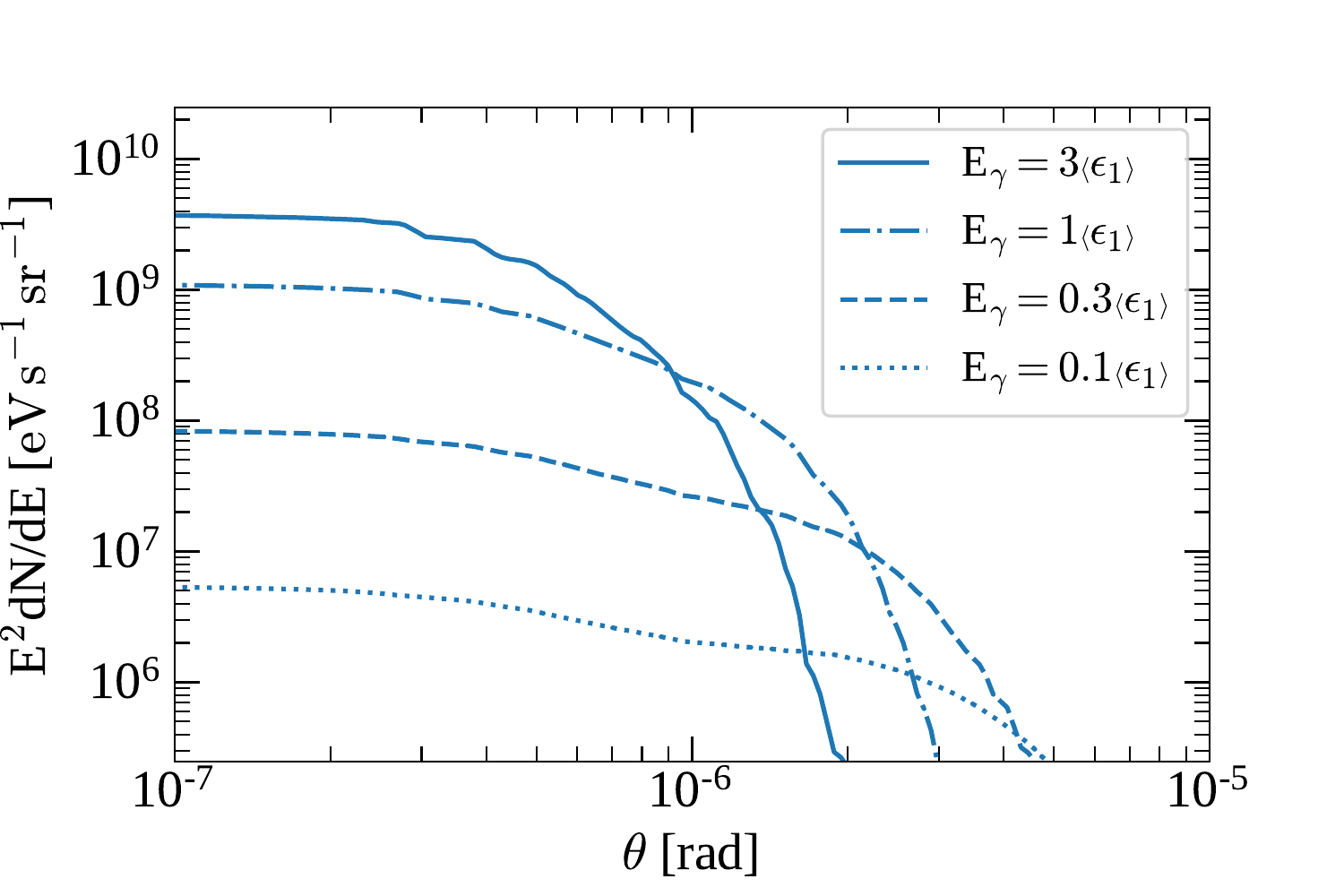}
  \caption{
    Inverse Compton emission profiles of a single electron with the Lorentz factor $\gamma=10^6$, scattering the isotropic black body photon field with the temperature $T=2.725$~K, evaluated at multiples of the mean scattered photon energy $\left< \epsilon_1 \right> = 3.6 \gamma^2 kT \approx 0.85$~GeV.
  }
  \label{fig::ic-angular-profile}
\end{figure}

\begin{figure}
  \includegraphics[width=\columnwidth]{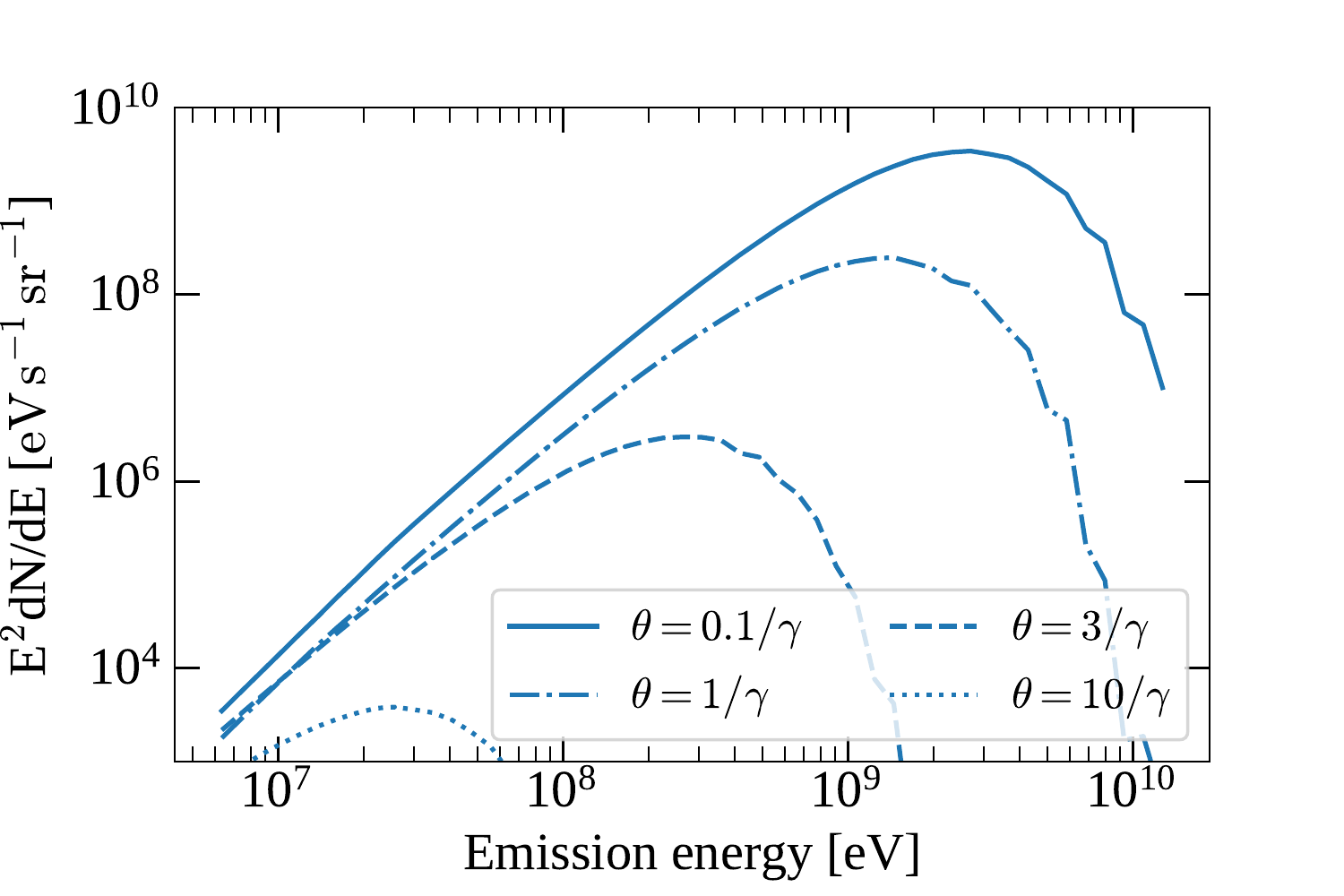}
  \caption{
    Inverse Compton emission spectra of a single electron with the Lorentz factor $\gamma=10^6$, scattering the isotropic black body photon field with the temperature $T=2.725$~K, evaluated at several offset angles with respect to the electron motion direction.
  }
  \label{fig::ic-angular-spectrum}
\end{figure}

The angular distributions of the photons resulting from the scattering a non-monochromatic background distribution can be found integrating Eq.~\ref{eq::ic-emissivity} over the corresponding radiation spectrum. For the specific case of the thermal background with the temperature $T=2.725$~K, corresponding to the cosmic microwave background (CMB) at redshift $z=0$, resulting distributions from an electron with the Lorentz factor $\gamma=10^6$ at several energies are shown in Fig.~\ref{fig::ic-angular-profile}. As expected, most of the emitted energy flux is concentrated within the narrow $1/\gamma$ cone. The emission spectra of the same electron in several directions are shown in Fig.~\ref{fig::ic-angular-spectrum}.

\subsection{Time delay from the geometrical path difference}

\begin{figure}
  \includegraphics[width=\columnwidth]{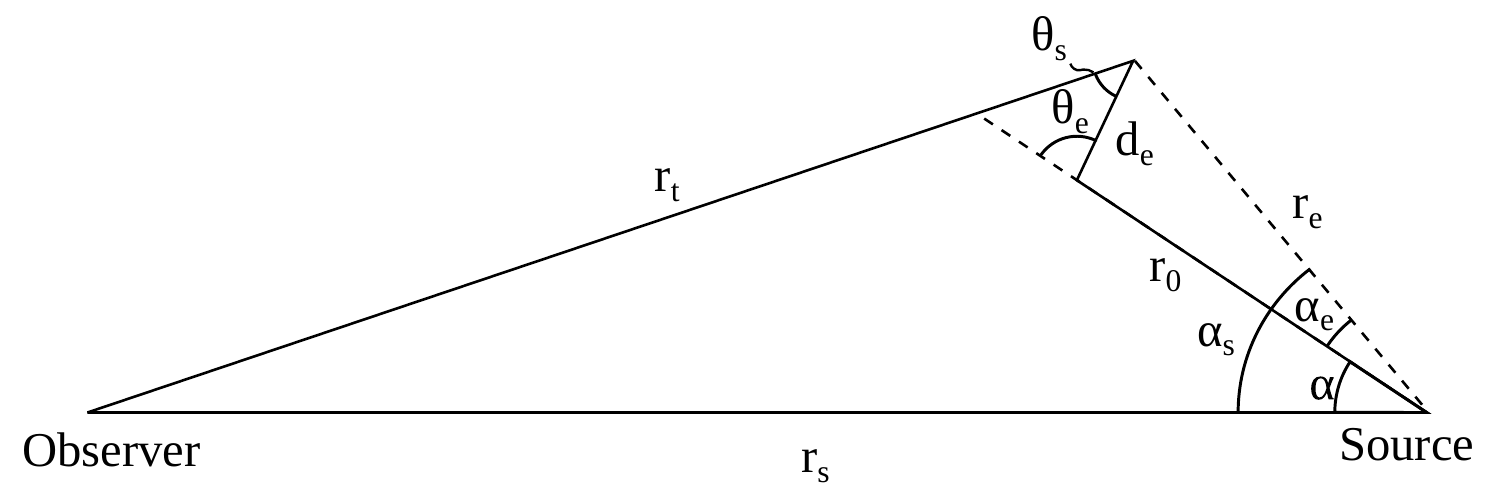}
  \caption{
    Sketch of the secondary emission problem. Observer (on the left) is separated from the source (on the right) by the distance $r_s$. An photon emitted by the source at an angle $\alpha$ is absorbed having travelled over the distance $r_0$ and generates an electron (positron) at the relative angle $\theta_e$. The latter travels over the distance $d_e$ before emitting the secondary photon reaching the observer after crossing the distance of $r_t$.
  }
  \label{fig::geometry}
\end{figure}

Geometry of the secondary emission time delay problem is depicted in Fig.~\ref{fig::geometry}. The resulting time delay may be written as a sum of the delays originating in the triangles formed by the sides $(r_0, d_e, r_e)$ on one hand side and $(r_s, r_e, r_t)$ on the other. For the first triangle the time delay is simply 

\begin{equation}
  \Delta t'_1 = t_e + (r_0 - r_e) / c \approx 
    t_e - \frac{d_e}{c} 
    \left(
      1 - \frac{r_0} {d_e + r_0} \frac{\theta_e^2}{2}
    \right)
\end{equation}
where $t_e$ is the exact time the electron (positron) required to travel over $d_e$ accounting for its gradual slowing down due to cooling.

For the second triangle the delay is
\begin{eqnarray}
  \Delta t'_2 = 
  \frac{r_s - r_e}{c}
  \left[
    \sqrt{1 + \frac {2 r_s r_e (1 - \cos \alpha_s)} {(r_s - r_e)^2}} - 1
  \right] \nonumber \\
  \approx
  \frac{1}{c}
  \frac{r_s r_e} {r_s - r_e} 
  \frac{\alpha_s^2} {2}
\end{eqnarray}
with $\alpha_s \approx \alpha^2 + \alpha_e^2 + 2 \alpha \alpha_e \cos \phi_e$, where $\phi_e$ is the positional angle of the electron (positron) motion direction and $\alpha_e \approx \frac{d_e}{r_e} \theta_e$.

The total time delay thus is
\begin{eqnarray}
  \Delta t = (1 + z) ( \Delta t'_1 + \Delta t'_2 ) \nonumber \\
    \approx (1 + z) 
    \left[
      t_e - \frac{d_e}{c} 
      \left(
        1 - \frac{r_0} {d_e + r_0} \frac{\theta_e^2}{2}
      \right)
    \right. \nonumber \\
    \left.
      + 
      \frac{1}{c}
      \frac{r_s r_e} {r_s - r_e} 
      \frac{\alpha_s^2} {2}
    \right]
    \label{eq::tdelay-total}
\end{eqnarray}
Neglecting the difference between $t_e$ and $\frac{d_e}{c}$ and assuming that the electron cooling distance is much smaller than the initial photon mean free path (i.e. $d_e \ll r_0$) while the source distance is much larger than it (i.e. $r_s \gg r_0 \sim r_e$), one finds
\begin{equation}
  \Delta t \approx (1 + z) \frac{1}{c}
    \left(
      d_e \frac{\theta_e^2}{2}
      + 
      r_0 \frac{\alpha_s^2} {2}
    \right)
\end{equation}
The suitable range of $\theta_e \lesssim 2 / \gamma$ here is defined by the electron/positron pair production angular scatter (see Sect.~\ref{sect::pp-angular}), whereas for $\alpha_s$ it is set by the requirement that the corresponding scattering angle with respect to the electron direction of motion is $\theta_{sc} \lesssim 1/\gamma$. This angle can be found from the same triangles
\begin{eqnarray}
  \theta_{sc}^2 \approx 
    \frac{r_s^2}{R^2} \alpha^2
    +
    \frac{2 r_s (r_s - r_0)}{R^2} \alpha \theta_e \cos \phi_e \nonumber \\
    +
    \frac{(r_s - r_0)^2}{R^2} \theta_e^2
    \label{eq::geom-scattering-angle}
\end{eqnarray}
where $R=r_s - (r_0 + d_e)$. It defines the scalar product ${\bf me}$, required to evaluate the inverse Compton emissivity in Eq.~\ref{eq::ic-emissivity}.

\subsection{Total light curve and spectral energy distribution}
\label{sect::total-lc-procedure}

To get the resulting light curve and spectrum of the secondary emission, the IC emissivity in Eq.~\ref{eq::ic-emissivity} needs to be integrated over the spatial and energy distribution of the generated electron-positron pairs. A Monte Carlo approach is used for this purpose, where $10^6$ source photons are generated in the energy range $0.01-10$~TeV. 
Since most of the astrophysical sources of gamma-ray emission are likely much smaller than the $D_\gamma \simeq 800 (E_\gamma/ 1~\text{TeV})^{-1}$~Mpc mean free path of the initial gamma rays~\cite{neronov_semikoz_09}, the emitting source is assumed to be point-like. The generated photons are taken uniformly distributed within the cone with half-opening angle $\alpha_{max}=10^{-5}$~rad, chosen to adequately cover the angular spread of the electrons with the Lorentz factor $\gamma \gtrsim 10^6$, mostly responsible for the emission in the target $0.1-1$~GeV energy range.

For each photon, an absorption probability on EBL is evaluated from Eq.~\ref{eq::ebl-probability-vs-redshift}. In case of absorption, the corresponding absorption redshift, EBL photon energy and interaction angle, randomly sampled from Eq.~\ref{eq::dtau-dz-ds}, are used to calculate the energy and propagation direction of the created electron and positron using Eqs.~\ref{eq::pp-dsigma-de} and~\ref{eq::pp-mu}. For each of those, the generated IC emission is calculated integrating Eq.~\ref{eq::ic-emissivity} over the isotropic thermal distribution of the CMB photons with the temperature $T = 2.725 (1 + z)$~K, where the scattering angle is defined by Eq.~\ref{eq::geom-scattering-angle}. At every point of their respective trajectories, the particle energies are adjusted according to the total energy loss due to IC emission. Finally, the corresponding emission time delay is evaluated using Eq.~\ref{eq::tdelay-total} for every point of the particle trajectory.

The obtained this way cascade emission kernel $K(\epsilon_1, E_\gamma, \Delta t)$ relates primary source emission at energy $E_\gamma$ to the generated secondary flux at energy $\epsilon_1$, arriving with the time delay $\Delta t$. The total secondary emission flux at an arbitrary moment $t$ is straightforwardly obtained as:
\begin{equation}
  F_{echo}(\epsilon_1, t) = 
    \int_{-inf}^{t} F(E_\gamma, t') K(\epsilon_1, E_\gamma, t-t') dt'
\end{equation}
where $F(E_\gamma, t')$ is the assumed intrinsic light curve of the source.

\begin{figure}
  \includegraphics[width=\columnwidth]{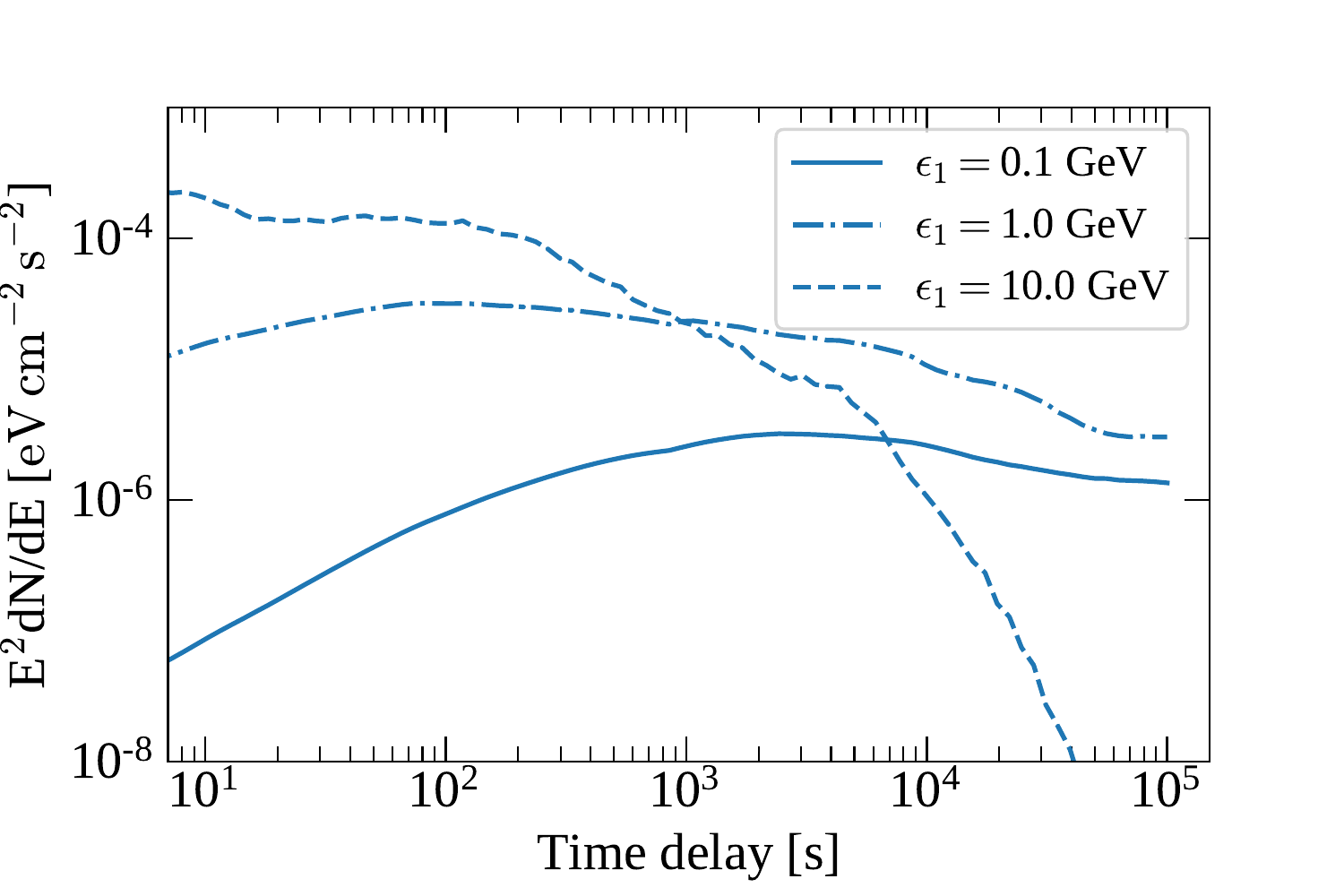}
  \caption{
    Time dependency of the cascade kernel $K(\epsilon_1, E_\gamma, \Delta t)$, integrated over the initial photon energy $E_\gamma$,  evaluated at several emission energies $\epsilon_1$. Calculation was performed for a putative source at the redshift of GRB190114C ($z=0.42$) with an exponentially cut off power law spectrum with the index $\Gamma = -2$, normalization at $E_0 = 100$~GeV of $N=10^{-22}~1/(\text{cm}^2 \text{ s eV})$ and the cut off energy $E_c = 5$~TeV.
  }
  \label{fig::cascade-kernel-lc}
\end{figure}

\begin{figure}
  \includegraphics[width=\columnwidth]{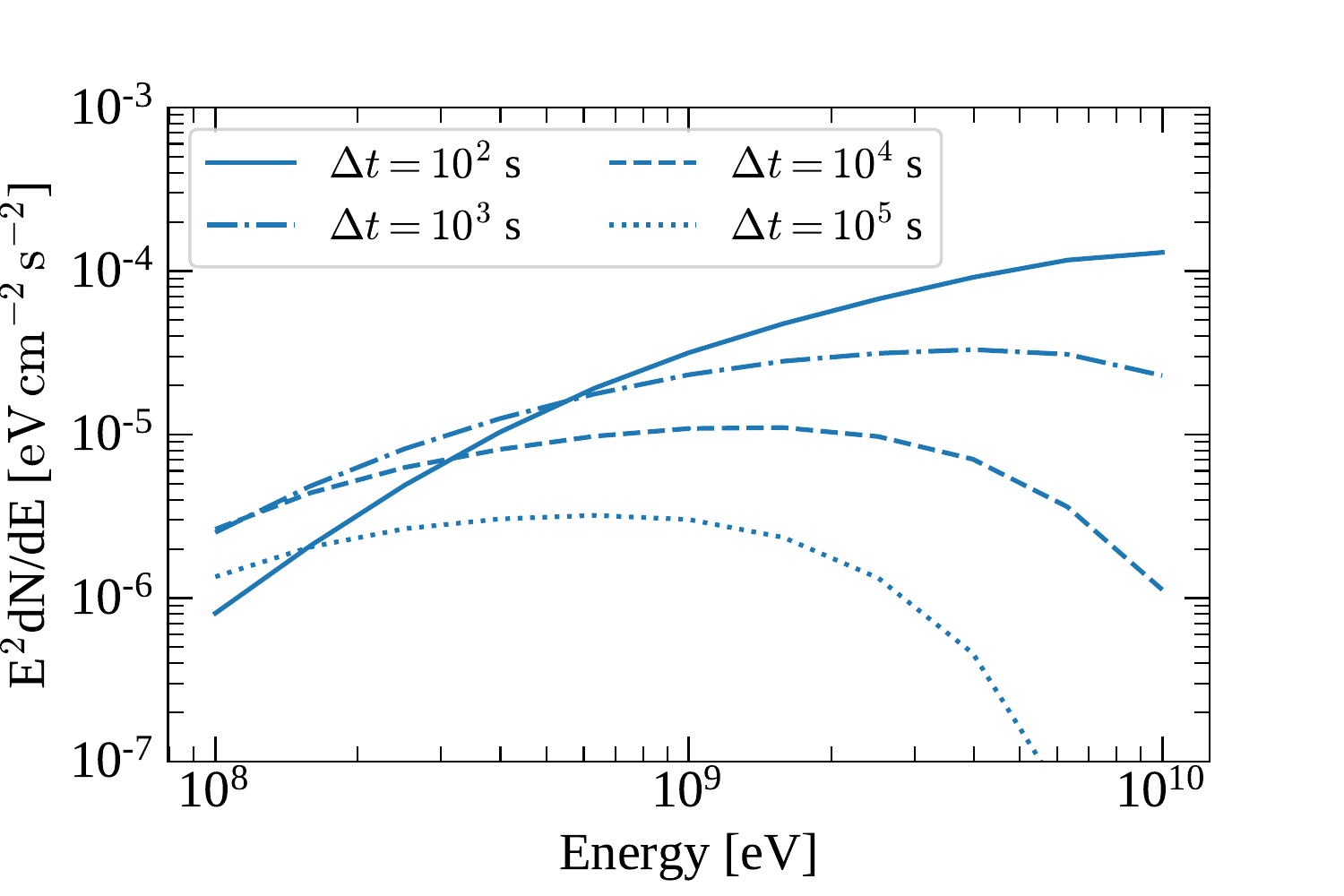}
  \caption{
    Energy dependency of the cascade kernel $K(\epsilon_1, E_\gamma, \Delta t)$, integrated over the initial photon energy $E_\gamma$ and evaluated at several values of time delay $\Delta t$. The assumed emission source is the same as in Fig.~\ref{fig::cascade-kernel-lc}.
  }
  \label{fig::cascade-kernel-spec}
\end{figure}

The energy and time dependence of cascade emission kernel is illustrated in Figs.~\ref{fig::cascade-kernel-lc} and~\ref{fig::cascade-kernel-spec} and can be understood from the relations outlines in Sect.~\ref{sect::pair-echo-calculation}. The smallest time delays correspond to the smallest electron direction offsets $\theta_e$, that during the pair production are achieved for the highest energy electrons~(see Fig.~\ref{fig::pp-rate-example}). This results in the hard emission spectrum peaking above few GeVs. With the increase of the time delay a larger part of the initial emission cone becomes visible -- though at a cost of a larger offset angle $\theta_e$ and / or scattering angle $\theta_{sc}$. As efficient IC emission takes place only within $\theta_{sc} \lesssim 1/\gamma$, the required increase in the offset angle $\theta_e$ lowers the maximal energy of the emitting particles, leading to an eventual gradual decrease of the generated energy flux starting from the highest energies.

\section{GRB190114C delayed emission modelling}

HE and VHE gamma-ray emission from GB190114C has been contemporaneously measured in the $0.1-1$~GeV and $0.3-1$~TeV energy ranges~\cite{magic_grb190114c_b}. Though it is possible that VHE emission is of the secondary origin, cascading down from $E \gtrsim 20$~TeV energies, the mean free path of such energetic primary photons with respect to pair production on EBL is at least 20 times shorter than that of the VHE photons in question~\citep{neronov_semikoz_09}. The additional time delay associated with such high energy secondaries does not exceed few seconds~\citep{takahashi08, neronov_semikoz_09}. The possible secondary origin of the VHE emission thus does not substantially modify the spatial structure of the cascade and has a minor impact on the secondary emission in the HE band, where time delays of the order of $\Delta t \sim 10^2 - 10^5$~s are expected (see Fig.~\ref{fig::cascade-kernel-lc}). Due to this an assumption, that the measured VHE emission represents the intrinsic source radiation, does not have a notable impact on the conclusions regarding the delayed emission in HE band. The model of the GRB190114C delayed emission used here thus constituted a prediction of the secondary flux in the $0.1-1$~GeV energy range given the measured flux in the $0.3-1$~TeV range as if it was of an intrinsic origin.

Worthy to note that, assuming that initial photon energy is distributed equally between the generated electron and positron, the energies of the primary $E_\gamma$ and secondary $\epsilon_1$ gamma-ray photons are related as $\epsilon_1 \approx 1 (E_{\gamma} / \text{1 TeV})^2$~GeV, implying that the HE and VHE windows used for GB190114C observations are well-suited for the secondary emission search.

\begin{figure}
  \includegraphics[width=\columnwidth]{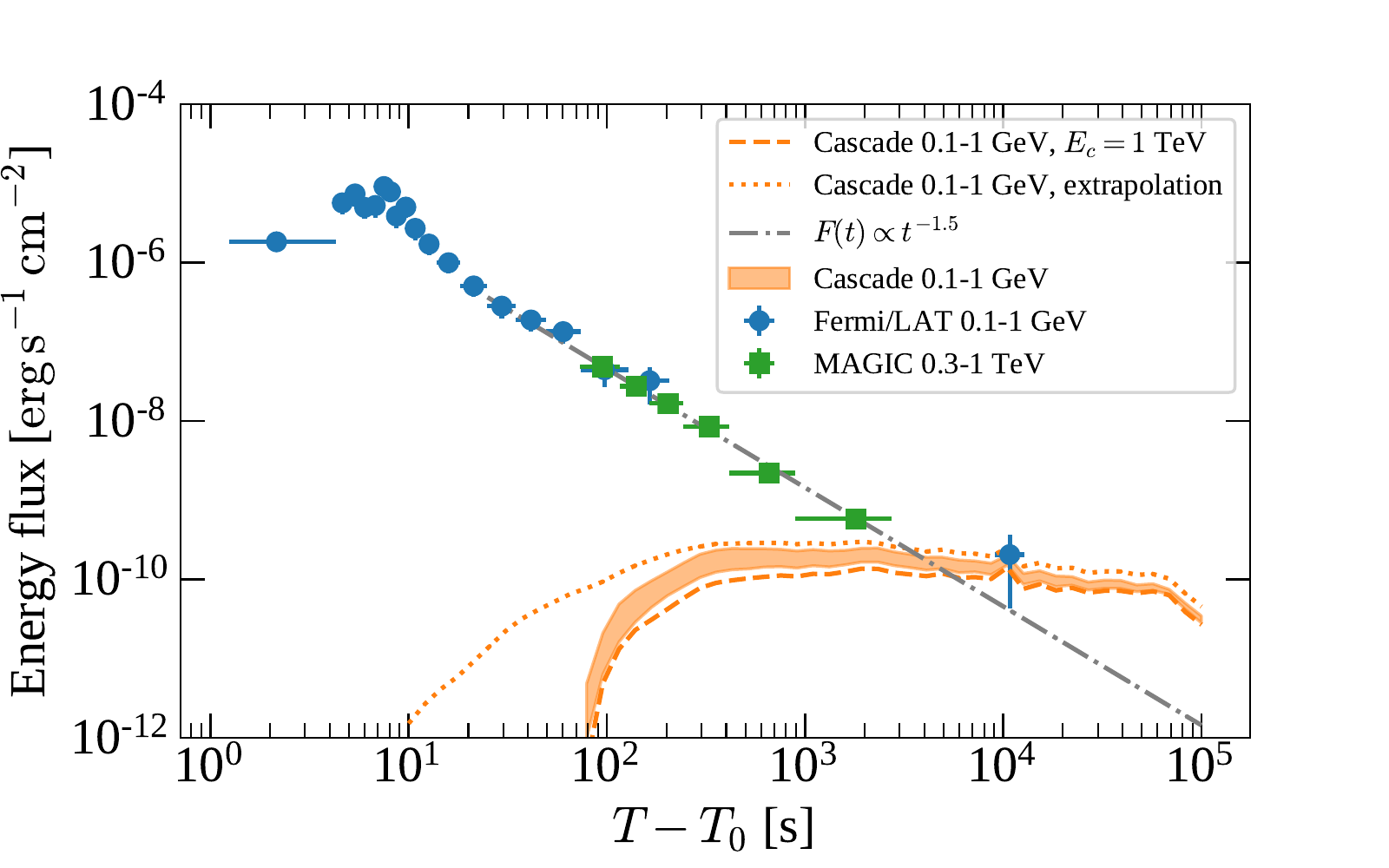}
  \caption{
    Expected secondary (cascade) flux from GRB190114C compared to the actual measurements in TeV and GeV bands~\cite{magic_grb190114c_b}. Cascade flux is estimated from the power law primary source emission in the $T-T_0=[68; 2400]$~s time window, where the VHE emission was measured. An estimate assuming an exponential energy cut-off at $E_c = 1$~TeV is shown with the dashed orange line. Extrapolation of the initial source flux assuming the $F(t) \propto t^{-1.5}$ scaling found in~\cite{magic_grb190114c_b} is shown with the gray dash-dotted line. Cascade resulting from this extrapolation down to $T-T_0 = 5$~s is depicted with the dotted orange line.
  }
  \label{fig::grb190114c-cascade-lc}
\end{figure}

Time-delayed cascade emission from GRB190114C was modelled using the procedure outlined in Sect.~\ref{sect::total-lc-procedure}. As the VHE spectrum of the source is consistent with the power law model, it was chosen as the intrinsic spectral shape in the model. The model parameters and their uncertainties were found from a $\chi^2$ fit to the reported spectral points in each of the 5 time bins spanning from $T=T_0+68$~s to $T=T_0+2400$~s, where such points were reported~\cite{magic_grb190114c_b}. The uncertainties were further propagated to the delayed flux estimate using a toy MC, where the cascade flux was re-calculated for 100 random realizations of the spectral parameters, sampled from the multi-variate normal distribution with the mean values and covariance matrix taken from the spectral fit. The resulting total time-delayed cascade flux is shown in Fig.~\ref{fig::grb190114c-cascade-lc} along with the corresponding HE/VHE measurements from~\cite{magic_grb190114c_b}.

Though the maximal energy of GRB190114C VHE emission is not constrained by VHE data, an artificial limit of the initial photon energy of $E_\gamma^{max} = 10$~TeV was assumed here for numerical reasons. At the same time, relation between the primary and secondary photon energies, mentioned above, suggest the bulk of the cascade emission, resulting from primary photons with $E_\gamma > 1$~TeV, will have energy $\epsilon_1 \gtrsim 1$~GeV and thus would not contribute to the flux in the $0.1-1$~GeV window considered here. To verify this assumption directly, the time-delayed cascade emission was re-calculated for a power law spectrum with a exponential cut-off at 1~TeV (also displayed in Fig.~\ref{fig::grb190114c-cascade-lc}). The resulting $\lesssim 30\%$ flux decrease at $T - T_0 \gtrsim 10^3$~s supports the assumption of the sub-dominant contribution of intrinsic $E_\gamma \gtrsim 1$~TeV emission to the predicted cascade.

It should be noted that the cascade flux estimate shown in Fig.~\ref{fig::grb190114c-cascade-lc} is conservative in the sense that it does not include any potential ``echo'' flux from $T-T_0 < 68$~s when MAGIC observations have started. Contribution of this early-time emission, however, is not decisive. Assuming the source flux evolves as $F(t) \propto t^{-\alpha}$ with $\alpha \approx 1.5-1.6$~\cite{magic_grb190114c_a, magic_grb190114c_b} while the spectral shape is the same as in the $T-T_0 = [68; 110]$~s interval (the earliest for which VHE measurement were published), relative contribution from interval $T-T_0 = [25; 68]$~s, starting at the approximate end of the GRB prompt emission phase, is around $\simeq 20$\% from the estimated cascade flux; extrapolation of the VHE emission to the prompt phase down to $T-T_0 = 5$~s leads to $\simeq 80$\% larger cascade flux, which is still within the uncertainties of the Fermi/LAT measurement.

\section{Discussion}

As one can see from Fig.~\ref{fig::grb190114c-cascade-lc}, the expected cascade flux from GRB190114C is compatible with the Fermi/LAT measurement at $T-T_0 \approx 10^4$~s, so that, in principle, the registered flux may be composed of the ``echo'' entirely. Though afterglow origin of this emission is possible~\citep{magic_grb190114c_b}, this demonstrates that in the absence of IGMF the cascade emission from bright GRBs similar to GRB190114C is detectable with the current generation of gamma-ray telescopes despite the strong dilution of the ``echo'' flux in time -- in contrast with earlier findings~\cite{dzhatdoev20}, focusing on the GRB190114C HE emission at later times $T-T_0 > 2\times 10^4$~s.

Temporal evolution of the pair echo flux, however, is distinctly different from that of the intrinsic GRB emission~(see Fig.~\ref{fig::grb190114c-cascade-lc}). This opens a possibility to distinguish the echo emission if the GRB light curve in the HE gamma-ray band would be measured for at least $10^4$~s without interruptions and / or theoretical arguments can be given for extrapolation of the intrinsic GRB flux from $T \approx T_0 + 10^2$~s on to later times. In a particular case of GRB190114C, extrapolation of the GeV band flux after the prompt GRB phase using the power law scaling $F(t) \propto t^{-\alpha}$ with $\alpha \approx 1.5-1.6$ measured in VHE band~\cite{magic_grb190114c_a, magic_grb190114c_b} suggests the intrinsic source flux is below the Fermi/LAT measurement, indicating the detection of the ``echo'' emission at $T-T_0 \approx 10^4$~s.

Such a detection would imply a low IGMF with $B \lesssim 10^{-21}$~G at the GRB190114C redshift of $z \approx 0.42$, due to the requirement that IGMF-induced deflections of the electron-positron pairs do not exceed their intrinsic scatter~\cite{neronov_semikoz_09}. With measurement using hard-spectrum gamma-ray loud AGNs suggesting $B \gtrsim 10^{-17}$~G at $z \sim 0.1$~\cite{neronov10, taylor11, tavecchio11, vovk12, fermi_igmf, HEGRA_IGMF, MAGIC_extended, HESS_IGMF, Veritas_IGMF}, such a detection would indicate a fast evolution of IGMF with redshift, thus strongly disfavoring the cosmological origin of IGMF, where the field strength scales with redshift as $B(z) \propto (1+z)^2$. It may, however, also present challenges for galactic-origin models of IGMF, where the field is expected to reach its present value around $z \sim 1$~\cite{garaldi21}. The latter tension may be, in principle, alleviated accounting for the highly inhomogeneous structure of IGMF, expected both for the primordial field, frozen into the plasma and following the matter density fluctuations~\cite[{e.g.}][]{blasi99}, and that originating from the galactic outflows~\cite{boyarsky_tng_bubbles_vs_seed} (though a small effect from the outflow-driven magnetic field bubbles was reported on average~\cite{bondarenko22}).

A detection of the secondary emission would be also indicative of the sub-dominant role of the plasma instabilities in cooling the injected electrons and positrons, that presently remains uncertain~\citep{broderick12, miniati13, chang14, shalaby18, vafin18, vafin19}.

It is interesting to note that if the detected emission at $T-T_0 \approx 10^4$~s indeed comes from the delayed cascade ``echo'', the VHE flux from GRB190114C during the prompt phase can not exceed much the prediction from the $F(t) \propto t^{-1.5}$ extrapolation to avoid tension with the measurements.

Worthy to note, that the GRB afterglow emission itself may contain time-delayed components mimicking the cascade ``echo''.
Structured GRB jets may result in flares and plateaus in the light curves, that are sometimes observed in X-ray band~\cite[{e.g.}][]{oganesyan20, salafia22};
the evolving ratio between the different spectral components of the GRB emission may lead to a change in the light curve slope, similar to that from the ``echo'' on-set~\citep{magic_grb190114c_b}.
Though temporal evolution of these phenomena in general differs from that of the ``echo'', the available data on GRB190114C are insufficient to tell them apart. If more than 80\% of the measured flux at this time is indeed due to the afterglow, the predicted cascade emission would be excluded, imposing a $B \gtrsim 10^{-21}$~G limit on IGMF strength at $z\approx 0.4$.

Clearly, a more careful assessment of the possible intrinsic GRB190114C emission contribution to the measured flux at $T-T_0 \approx 10^4$~s is required to assess the reliability of the ``echo'' emission detection assumption.

Such an ``echo'' emission can be as well searched for from other flaring gamma-ray sources at $z \gtrsim 0.1$.
Several other GRBs have also been detected in VHE band at redshifts ranging from $z \approx 0.08$ to 1.1~\cite{hess_grb180720b, hess_grb190829a, magic_grb201216c_gcn}. With the measured VHE fluxes much lower than in the  case of GRB190114C, the expected pair echo flux for these GRBs would fall below the Fermi/LAT detection limit. However these detections demonstrate that a substantial number of GRBs may feature hard VHE spectra with the power law index $\Gamma \gtrsim -2$ and emission longer than $10^2-10^3$~s, required for a detectable pair echo in the absence of IGMF. The emerging population of such sources may be crucial for IGMF measurements at redshift $z\gtrsim1$.


\begin{acknowledgments}
  Author gratefully acknowledges the support of the Institute for Cosmic Ray Research (ICRR), the University of Tokyo in realization of this study and that of the CTA-North computing center at La Palma, Spain, for providing the necessary computational resources.
\end{acknowledgments}


\bibliographystyle{myrevd}
\bibliography{references}

\end{document}